\documentclass[aps,prd,reprint,nofootinbib, superscriptaddress]{revtex4-1}

\usepackage{amsmath,amssymb, amsthm,amstext}
\usepackage{natbib}
\usepackage{graphicx}
\usepackage{color}
\usepackage{array, enumerate}
\usepackage{bm}
\usepackage{multirow}
\usepackage[breaklinks,colorlinks,citecolor=blue]{hyperref}

\def\be{\begin{equation}}
\def\ee{\end{equation}}

\newcommand{\rs}{r_*}
\newcommand{\Ric}{{\mathcal R}_{\rm IC}}
\newcommand{\Rgg}{{\mathcal R}_{\rm pair}}
\newcommand{\pold}{p^{\rm old}}
\newcommand{\BNS}{B_{\rm NS}}
\newcommand{\rNS}{r_{\rm NS}}

\begin{document}
\title{Black Hole Discharge: \\ very-high-energy gamma rays from black hole-neutron star mergers}
\author{Zhen Pan}
\email{zpan@perimeterinstitute.ca}
\affiliation{Perimeter Institute for Theoretical Physics, ON, N2L2Y5, Canada}
\author{Huan Yang}
\email{hyang@perimeterinstitute.ca}
\affiliation{Perimeter Institute for Theoretical Physics, ON, N2L2Y5, Canada}
\affiliation{University of Guelph, Guelph, Ontario N2L 3G1, Canada}

\date{\today}
\begin{abstract}
 With mass ratio larger than $\sim 5$ (which depends on the black hole spin and the star radius),  star disruption is not expected for a  black hole  merging with a neutron star  during the final plunge phase.
In the late inspiral stage, the  black hole is likely  charged as it cuts through the magnetic field carried by the neutron star, leaving a temporarily charged black hole after merger.
The unstable charged state of the remnant black hole rapidly  neutralizes by interacting with the surrounding plasma and photons,
which we investigate in first principle by numerically solving a coupled set of Boltzmann equations  of 1+1 form for
non-spinning BH background.
The resulting basic picture is as follows. Electrons and positrons are accelerated in the BH electric field,
which then lose energy to surrounding soft photons via Compton scattering;
more electrons and positrons will be created from pair production as the hard photons colliding with soft photons, or through the Schwinger  process
in strong electromagnetic fields. The cascade stops  when the charged black hole accretes enough opposite charges and
becomes neutralized. We find that $\sim 10\%$ (which depends on the soft photon energy and number density)
of the total electric energy is
carried away to infinity in a time interval $\sim 1$ ms by very-high-energy ($>50$ GeV, the low energy detection threshold of the MAGIC telescope) gamma rays whose  spectrum is approximately a power law with spectral index $\sim -2.3$.  We expect the discharge picture to be true for spinning charged BHs as well.
\end{abstract}
\maketitle

\section{Introduction}
Short gamma-ray burst (sGRB) has long been suggested to be associated with double neutron star (NS) mergers - a conjecture lately supported by the multi-messenger
detection of gravitational waves (GWs) and various electromagnetic (EM) counterparts in GW170817 \cite{LIGO2017a,LIGO2017b}.
Another possible origin of sGRBs is merging black hole (BH) and NS binary.
If the BH disrupts the NS during the late inspiral phase, the remnant matter may form a temporary accretion disk around the final BH and power an energetic jet along with gamma ray emission \cite{Janka1999}. This scenario may be tested with imminent detection of BH/NS merger events in the O3 run of LIGO/Virgo collaboration, together with EM observations.  Future third-generation GW detectors may be able to probe the NS disruption \cite{Miao:2017qot,martynov2019exploring} and help distinguish possible low-mass BH/NS binaries from binary NSs \cite{Yang:2017gfb,Hinderer:2018pei,Chen:2019aiw}.
On the other hand, if the BH/NS binary mass ratio is beyond certain threshold  $\sim 5$ which depends on the BH spin
and the NS radius  (\cite{Shibata2009,Etienne2009,Kyutoku2011,Foucart2012}, dubbed as type III merger therein and hereafter), no star disruption is expected. The BH basically swallows the NS, and the post-merger waveform is dominated by BH ringdowns \cite{Mark:2014aja,Dias:2015wqa}. The associated electromagnetic emission during the merger mainly originates from the surrounding plasma within the system, which has a much smaller energy reservoir than a typical sGRB.
A few mechanisms that might lighten up these systems have been proposed,
e.g., DC circuit \cite{McWilliams2011, Lai2012}, BH pulsar \cite{Levin2018a}, BH electric generator \cite{Dai2019} and electric/magnetic dipole radiation \cite{zhang2016mergers,Zhang2019}.
Previous discussions focused on estimating  the energy budget rather than description about the plasma dynamics and emission properties, so that  the spectroscopic prediction
of EM counterparts has not been made.

\begin{figure}
\includegraphics[scale=0.28]{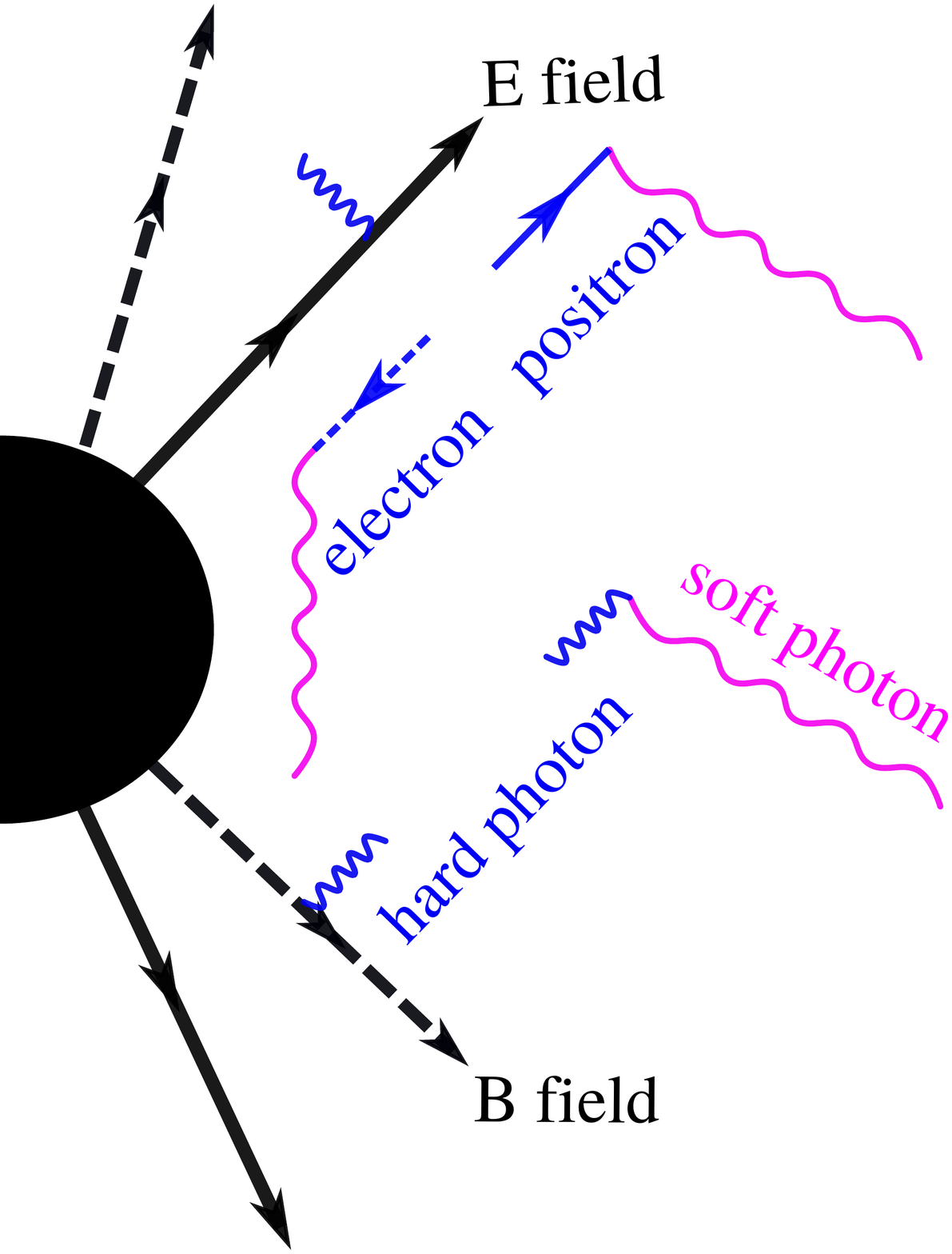}
\caption{\label{fig:cartoon}
A cartoon picture summarizing  major reactions in our model:
electrons and positrons are accelerated in opposite directions by the electric field;
soft photons gain energy via IC scatterings with electrons and positrons;
$e^\pm$ pairs are created as hard photons colliding with background soft photons, or through Schwinger process in electromagnetic fields.
 Vacuum polarization in charged BHs was originally considered in \cite{Gibbons:1975kk}.}
\end{figure}

In this paper, we report a first-principle study of EM signals from the aftermath of type III BH/NS mergers.
 Considering a typical NS mass  $\sim 1.4 M_\odot$, the corresponding BH mass threshold is already near the lower end of the observed BH mass in X-ray binaries \cite{tetarenko2016watchdog,Strader:2012wj,chomiuk2013radio,Miller-Jones:2015zba}.
To simplify the set-up, we  neglect the BH spin and assume the final BH to be Schwarzschild.
The amount of electric charge carried by the BH can be roughly estimated as (also see \cite{Levin2018a})
$Q\approx E M^2 \approx \BNS\rNS^2 $, where $M$ is BH mass and $E\sim  (v_{\rm rel}/c) \BNS \approx \BNS$  is the magnitude of the induced electric field.\footnote{As shown in \cite{Zhang2019}, the NS also carries net charge
$\sim \Omega_{\rm NS} B_{\rm NS} r_{\rm NS}^3$, which is smaller than the BH charge by a factor $\sim \Omega_{\rm NS} r_{\rm NS}$, where $\Omega_{\rm NS}$ is the angular velocity of the NS.}  To get an intuition about the BH charge $Q$, we define a dimensionless quantity
\be \Gamma_Q \equiv \frac{Qe}{Mm_e} \ , \ee
which is the ratio of electric force and gravitational force on an electron outside the charge BH.
For $M = 10 M_\odot$, $\BNS = 10^{12}$ G and $\rNS = 10$ km, we have
$\Gamma_{Q,12} \approx 4\times10^{14}$ and total electric energy $Q_{,12}^2/4M \approx 10^{42} {\rm ergs}$,
where the lower index $_{12}$ is the power index of $\BNS \, [\rm G]$.
Electrons and positrons are accelerated in the BH electric field,
which consequently produce hard photons via inverse Compton (IC) scattering with surrounding soft photons.
More electrons and positrons will be generated as a result of pair production of the hard photons interacting with soft photons
and electromagnetic fields (see Fig.~\ref{fig:cartoon}). The above cascade stops only when the charged BH  neutralizes by accreting enough opposite charges.
As we will show later $\sim 10\%$ of the total energy is carries away by $\sim 10^{41}$ gamma-ray photons with
energy $> 50$ GeV, which is the low-energy detection threshold of the MAGIC telescope \cite{magic2016}.

We use geometric units $G=c=1$ and the Schwarzschild metric is written as
\be
ds^2 = -\alpha^2 dt^2 + \alpha^{-2} dr^2 + r^2d\Omega^2 \ ,
\ee
where $\alpha := \sqrt{1-2M/r}$ and we assume $M = 10 M_\odot$ in this paper.
Tortoise coordinate $\rs \equiv r + 2M \ln |\frac{r}{2M}-1|$ is used in our numerical simulations.

\section{Governing Equations}
The single particle distribution function $f(t, r, p)$ in phase space is defined
by counting the particle number $dN$ within the volume $(r, r+dr)$ and  $(p, p+dp)$:
$dN =   f(t, r, p) 4\pi \alpha^{-2} drdp = f(t, r, p) 4\pi d\rs dp $, where $p$ is the contravariant component
of momentum in the $r$ direction $p \equiv p^r = \sqrt{(p_t)^2-\alpha^2 m^2}$, with $-p_t$ being the
particle energy measured by observers at infinity and $m$ being the particle mass.
Boltzmann equation in the covariant form is written as \cite[e.g.,][]{Mezzacappa1989}
\be
p^\mu \frac{\partial f}{\partial x^\mu} - \Gamma^i_{\mu\nu}p^\mu p^\nu \frac{\partial f}{\partial p^i} + eF^{\mu\nu}p_\nu\frac{\partial f}{\partial p^\mu} = C[f] \ ,
\ee
where $\Gamma^i_{\mu\nu}$ is the metric connection; $F^{\mu\nu}$ is the Maxwell tensor; $\mu,\nu$ are spacetime indices and $i$ is a spatial index.
For the system of spherical symmetry we are considering, the Boltzmann equations for distribution functions
of electrons $f_+(t, r, p_+)$, positrons $f_-(t, r, p_-)$ and hard photons $f_\gamma(t, r, p_\gamma)$ are written as
\be
\label{eq:Boltzmann}
\begin{aligned}
\partial_t f_+(p_+) + v(p_+) \partial_{\rs} f_+(p_+)
+ a_+\partial_p f_+(p_+)
&= C_+ \ , \\
\partial_t f_-(p_-) + v(p_-) \partial_{\rs} f_-(p_-)
+ a_- \partial_p f_-(p_-)
&= C_- \ , \\
\partial_t f_\gamma(p_\gamma) +  v(p_\gamma) \partial_{\rs} f_\gamma(p_\gamma)
&= C_\gamma  \ ,
\end{aligned}
\ee
where $v(p) = p/(-p_t)$; $a_\pm = \pm \alpha^2 N_Q e^2/r^2 m_e$ are the acceleration of positrons and electrons
in the electric field, with $N_Q(r)e$ being the total charge enclosed by radius $r$.  Here we have ignored the gravitational force term which is much smaller than the electric force term. The collision terms $C_{\pm,\gamma}$ are contributed by IC scatterings and pair productions (see Fig.~\ref{fig:cartoon}),
\begin{subequations}
\begin{align}
C_+=&
-\Ric(p_+) f_+(p_+) + \Ric(\pold_+)  f_+(\pold_+)\frac{d\pold_+}{dp_+} \nonumber \\
&+\Rgg(\pold_\gamma) f_\gamma(\pold_\gamma)\frac{d\pold_\gamma}{dp_+} \ , \label{eq:Ca}\\
C_-=&
-\Ric(p_-) f_-(p_-) + \Ric(\pold_-)  f_-(\pold_-)\frac{d\pold_-}{dp_-} \nonumber \\
&+\Rgg(\pold_\gamma) f_\gamma(\pold_\gamma)\frac{d\pold_\gamma}{dp_-} \ , \label{eq:Cb} \\
C_\gamma =&
-\Rgg(p_\gamma) f_\gamma(p_\gamma) + \Ric(\pold_+) f_+(\pold_+)\frac{d\pold_+}{dp_\gamma} \nonumber \\
&+\Ric(\pold_-) f_-(\pold_-)\frac{d\pold_-}{dp_\gamma} \ , \label{eq:Cc}
\end{align}
\end{subequations}
where
$\Ric$  is the Compton scattering rate of a lepton in the soft photon background, and $\Rgg$ is the total pair production rate of a hard photon colliding with soft photons and through Schwinger process in strong electromagnetic fields ($\gamma^{\rm hard} + \gamma^{\rm soft}/E/B \rightarrow e^+ + e^-$), i.e., $\Rgg = \mathcal R_{\gamma\gamma} + \mathcal R_{\gamma E}+\mathcal R_{\gamma B}$. On the right hand side of Eq.~(\ref{eq:Ca}), $\pold_+$ is the momentum of positrons that slow down
to $p_+$ after one IC scattering; $\pold_\gamma$ is the momentum of hard photons
that annihilate into an $e^\pm$ pairs with momentum $p_{+}$, i.e.,
\be
\begin{aligned}
    p_+^{\rm old}    &=  \frac{p_+}{1- \alpha^{-2} E_\gamma^{\rm soft} p_+/m_e^2}& \ &({\rm for\ Inverse\ Compton}) ,\\
    p_\gamma^{\rm old} &= 2p_+ = 2p_-&\ &({\rm for\ pair\ production}),
\end{aligned}
\ee
where $E_\gamma^{\rm soft}:= -p_{t,\gamma}^{\rm soft}$ is the soft photon energy measured by observers at infinity, and the
factor $\alpha^{-2}$ takes account of the gravitational redshift in the vicinity of the BH.
In Eqs.~(\ref{eq:Cb},\ref{eq:Cc}), $\pold_{\pm,\gamma}$ are defined in a similar way.

To close the Boltzmann equations (\ref{eq:Boltzmann}), we also need the sourced Maxwell's equations,
which are written in terms
of $N_Q$ and $f_\pm(p)$ as follow,
\begin{subequations}
\begin{align}
    \partial_t N_Q   &= -4\pi \int_{-\infty}^{\infty}   v(p)\left[f_+(p) - f_-(p) \right]  dp\ , \label{eq:Maxwella}\\
    \partial_{\rs} N_Q &= 4 \pi \int_{-\infty}^{\infty} \left[ f_+(p) - f_-(p) \right]  dp \ , \label{eq:Maxwellb}
\end{align}
\end{subequations}
where Eq.~(\ref{eq:Maxwellb}) is a constraint of charge conservation, instead of an evolution equation.
To ensure the charge conservation during evolution, we in fact evolve two extended Maxwell's equations
in simulations (see Ref.~\cite{Palenzuela2010} and references therein for more details).

\section{Model Setup}
We take the moment of BH-NS merger as our starting point $t=0$, when we expect a charged BH and a remnant NS
magnetosphere, as well as some charged particles outside the BH. In this paper, we focus on investigating the BH discharge
process, with simplified assumptions of other relevant processes. Here we do not do a full Boltzmann+Maxwell analysis for the
remnant NS magnetosphere, instead we assume the remnant NS magnetopshere
is detached from the BH and expands outward after the merger.\footnote{The fate of
the remnant NS magnetosphere is still an open question now.
We expect an isolated BH cannot sustain a magnetosphere according to the ``no-hair" theorem.
There was some counterargument (e.g., in Ref.~\cite{Lyutikov2011}) claiming
that the ``no-hair" theorem does not apply here because they found an isolated BH can sustain the captured magnetic field for a long time \emph{assuming} the BH is able to self-produce a highly conducting plasma and maintain a high magnetization. It is not clear whether the crucial assumption adopted in their simulation, i.e., dissipationless ideal MHD and force-free plasma, is physically realistic as dissipative plasma processes such as pair creation are evitable in this scenario. Slight different assumptions lead to completely different conclusions even for the
non-spinning BH case \cite{Lyutikov2011,Nathanail:2017wly}.

Even the claim that the BH can sustain the captured magnetic field for a long time was true, the net BH charge in such state is about
$0.01 a^3 B_{\rm NS} r_{\rm NS}^2$ as estimated in Ref.~\cite{Levin2018a}, which is much smaller than the amount the BH acquires in the merger.  Here $a$ is
the dimensionless BH spin. Therefore, BH discharge process is inevitably in either scenario.}

The initial distribution functions $f_{\pm,\gamma}(r,p)$,
electric charge $Q$ carried by the BH  (or equivalently the dimensionless quantity $\Gamma_Q$)
and all the reaction coefficients: $\Ric, \mathcal R_{\gamma\gamma}, \mathcal R_{\gamma E}, \mathcal R_{\gamma B}$ are specified as follows.

\subsection{remnant magnetopshere and initial conditions}
Motivated by the dipole geometry  of magnetic field lines carried by the NS, we assume an initial magnetic field  of the remnant NS magnetosphere
\be \label{eq:B_t0}
B(r)|_{t=0} = \BNS(r_{\rm NS}/r)^3 \qquad ({\rm for}\ r > r_c = 6M) \,.
\ee
The  soft photon energy and number density are estimated assuming they mainly come
from cyclotron radiation by the non-relativistic leptons (with Goldreich-Julian density  \cite{Goldreich1969a})
in the magnetic field, i.e.,
\be\label{eq:soft}
\begin{aligned}
  E_\gamma^{\rm soft} &= \hbar \omega_B = \hbar\frac{eB}{m_e}\ , \\
  n_\gamma^{\rm soft}(r) &\approx \int_{r_c}^\infty n_{\rm GJ}(r') f_{\rm cyc}(r') \frac{r'^2}{|\vec r'-\vec r|^2} dr'\ ,
\end{aligned}
\ee
where $n_{\rm GJ}= 0.07(B/{\rm G})\ {\rm cm}^{-3}$ is the Goldreich-Julian number density (for a NS magnetosphere with rotation period of $1$ sec), $ f_{\rm cyc} \approx \frac{2}{3}\frac{e^2}{\hbar c}\omega_B$ is the number of soft photons emitted per unit time from a non-relativistic lepton moving in the magnetic field.

The energy of
soft photons varies from  maximum value $\sim 10$ eV to zero depending on the magnitude of underlying magnetic field. For our purpose here, only the soft
photons at the high-energy end affect both  emission and absorption of hard photons. Therefore,
we assume all soft photons are of the same energy $E_\gamma^{\rm soft}\in[1, 10]$ eV.

For a given $\Gamma_Q$, we set $f_+|_{t=0}=f_\gamma|_{t=0}=0$ and
\[ f_-|_{t=0} dp = \frac{n(r)}{\sqrt{2\pi}\sigma_p}\exp\left(-\frac{(p/\Gamma_Q m_e)^2}{2\sigma_p^2} \right)
d\left(\frac{p}{\Gamma_Q m_e}\right), \]
with $\sigma_p = 0.1$ and $n(r)/r^2 = c_n /(\exp((r_*-6M)/2M)+1) $,
where $c_n$ is some normalization constant ensuring
electric neutrality at infinity. As we will show later, the discharge process
and final products mainly depend on the initial amount of total electric energy, while
the initial distribution function $f_-$ does not affect the evolution much as long as
the initial kinetic energy of electrons is a minor component.

\subsection{reaction coefficients}
As a result of Eq.~(\ref{eq:soft}), we can parameterize the rate of a lepton scattered off soft photons as
\be
\Ric = n_\gamma^{\rm soft} \sigma_{\rm T}  =  (10 N_\gamma^{\rm soft})  \times \alpha^2r_c^5 /(r_c^5 + r^5)  M^{-1},
\ee
where $\sigma_{\rm T}$ is the Thomson cross section. $N_\gamma^{\rm soft}$ is a free parameter of $O(1)$ and the factor $\alpha^2$ is included to mimic the lower photon number density in the vicinity of the BH due to BH capturing.

For the soft photon background above, the coefficient  $\mathcal R_{\gamma\gamma} $ is written as
$\mathcal R_{\gamma\gamma} =  \xi_{\gamma\gamma}(p, E_\gamma^{\rm soft}) (\sigma_{\gamma\gamma}/\sigma_{\rm T})\Ric$,
where $\sigma_{\gamma\gamma}\approx 0.35\sigma_{\rm T}$ is the cross section of pair creation $\gamma+\gamma\rightarrow e^++e^-$
\cite{Gould1967} and factor $\xi_{\gamma\gamma}$ takes account of the energy threshold below which this process is prohibited. The pair creation coefficients $\mathcal R_{\gamma E}$ and $\mathcal R_{\gamma B}$
are known as \cite[e.g.,][]{Daugherty1983}
\be
\mathcal R_{\gamma X} = \frac{0.23}{a_0} X' \exp\left(- \frac{4}{3xX'}\right) \ ,
\ee
where $X=\{E,B\}$,  $a_0$ is the Bohr radius, $x = p_\gamma/2m_e$, $X' = X/X_{\rm cr}$ and $X_{\rm cr} \equiv m^2c^3/\hbar e = 4.4\times 10^{13}$ G. In our model, we obtain  electric field $E(t,r)$  self consistently
and assume the remnant magnetosphere expands outward in the light speed which determines the evolution of magnetic field $B(t,r)$ as
\footnote{While the exact expansion speed is unknown, the propagation velocities
of characteristic electromagnetic waves in the remnant magnetosphere are close to light speed. Considering different directions of these waves move in, we expect the remnant magnetosphere expand in a sub-light speed, whose exact value barely affects the discharge process.
\be
B(t,r) = B_{\rm NS} \frac{r_{\rm NS}^3}{(r-t)^3} \left(\frac{r-t}{r}\right)^{4/3}\ ({\rm for}\ r-t > r_c),
\ee}

For notation simplicity, we define a two-element array $\mathcal R_{\gamma\#}:= \{\mathcal R_{\gamma E}|_{E= (N_Qe/r^2)(\Gamma_{Q,\#}/\Gamma_Q) }, \mathcal R_{\gamma B}|_{\BNS=10^{\#} {\rm G}}\}$ for cases with $\BNS=10^{\#}$ G, e.g.,
$\mathcal R_{\gamma12}$ is a short-hand notation for $\{\mathcal R_{\gamma E}|_{E= (N_Qe/r^2)(\Gamma_{Q,12}/\Gamma_Q) }, \mathcal R_{\gamma B}|_{\BNS=10^{12} {\rm G}}\}$ for cases with $\BNS = 10^{12} {\rm G}$.

\subsection{implicit assumptions}
A few extra assumptions have been made in the model setup and we summarize them as follows.

Compton scatterings of leptons with hard photons are not included considering the lower
number density of hard photons compared with that of soft photons and the smaller cross section which is suppressed by
a factor $\sim p_\pm p_\gamma/m_e^2$ in high-energy limit.

We do not track the evolution of soft photons in the Boltzmann equations, instead we approximate them as a background field
with number density $n_\gamma^{\rm soft}(r)$ solely determined by cyclotron radiation from the remnant magnetosphere, since
the optical depth of soft photons due to collisions with high-energy pairs and hard photons $\sim n_{\pm,\gamma}\sigma_{\rm T} M$
is much smaller than unity for the problem we are considering.

The detailed evolution of the remnant magnetosphere is another controversial problem even for non-rotating remnant BHs \cite[e.g.,][]{Lyutikov2011,Nathanail:2017wly}. In spite of these certainties, one certain thing is that the remnant magnetopshere
is magnetically dominant, $B= B_r\gg E_\perp$, where $E_\perp$ is the perpendicular E-field component ensuring another Maxwell's equation
$\partial_t \vec B_r = -\nabla\times \vec E_\perp$. In this paper, we have ignored the small component $E_\perp$, and only keep $E_r$ and $B_r$.

\section{BH discharge and very-high-energy gamma rays}
\subsection{numerical setup}

To evolve the governing equations (\ref{eq:Boltzmann},\ref{eq:Maxwella},\ref{eq:Maxwellb}),
we use the 4-th order Runge-Kutta method on a $512\times 128$ grid (default resolution) covering
the phase space of $[r_{*,\rm min}, r_{*,\rm max}]\times \left([-p_{\rm max}, -p_{\rm min}]\bigoplus [p_{\rm min}, p_{\rm max}]\right)$, where we take
$r_{*,\rm min} = -8.6 M, r_{*,\rm max} = 56.5 M$ (corresponding to $r_{\rm min} = 2.01 M, r_{\rm max} = 50 M$),
and $p_{\rm max} = \Gamma_Q m_e, p_{\rm min} \lesssim 0.01 m_e^2/E_\gamma^{\rm soft}$.

We use the ingoing boundary condition at ``horizon" $r=r_{*,\rm min}$ and the outgoing boundary condition
at ``infinity" $r = r_{*,\rm max}$, i.e., no particle enters the computation domain from two boundaries.
For the boundaries in the $p$ direction, we naturally choose $f_{\pm,\gamma}(p=\pm p_{\rm max}) = 0$, i.e., no particle
is accelerated to the maximum energy $\Gamma_Q m_e$.

The convergence test and conservation test for our code have been done for a fiducial model and  are shown in the Appendix.

\subsection{simulation results}
\begin{figure*}
\includegraphics[scale=0.6]{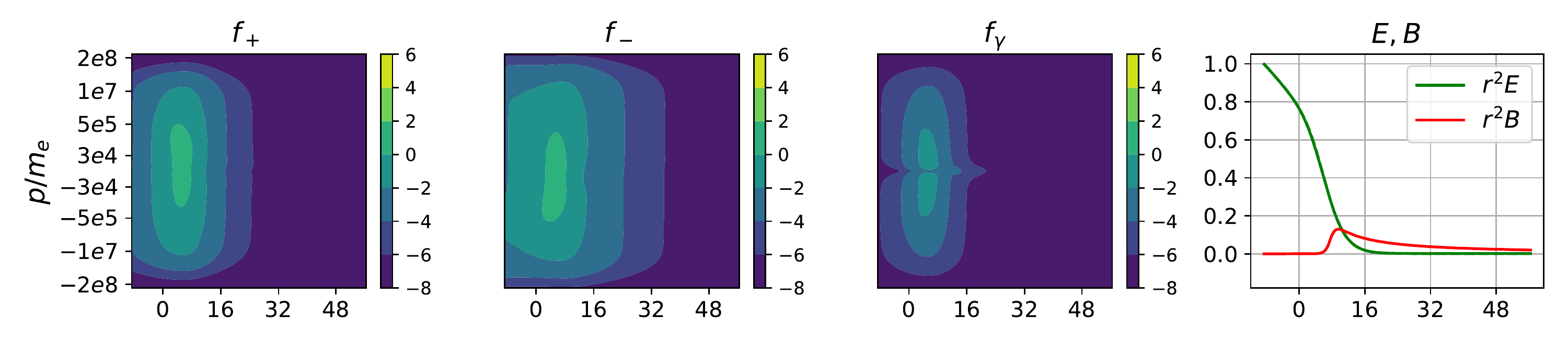}\\
\includegraphics[scale=0.6]{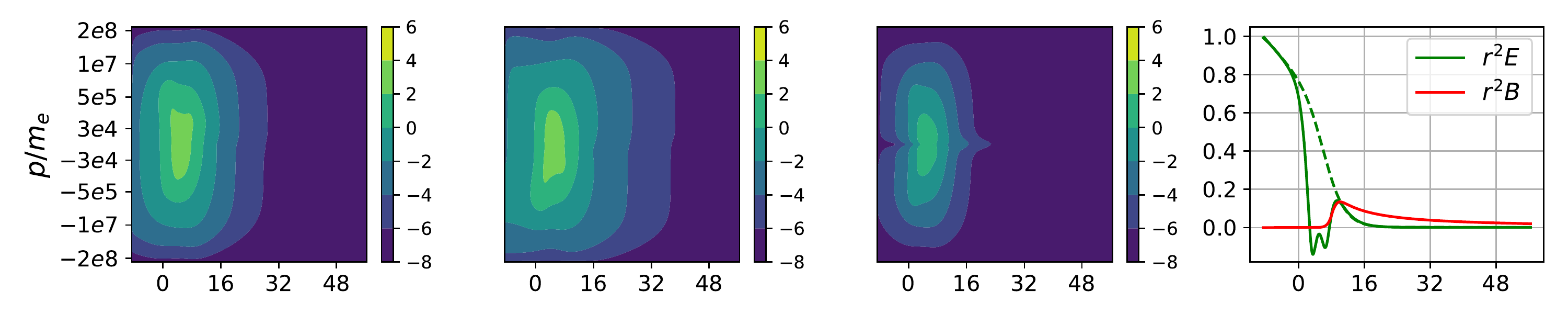} \\
\includegraphics[scale=0.6]{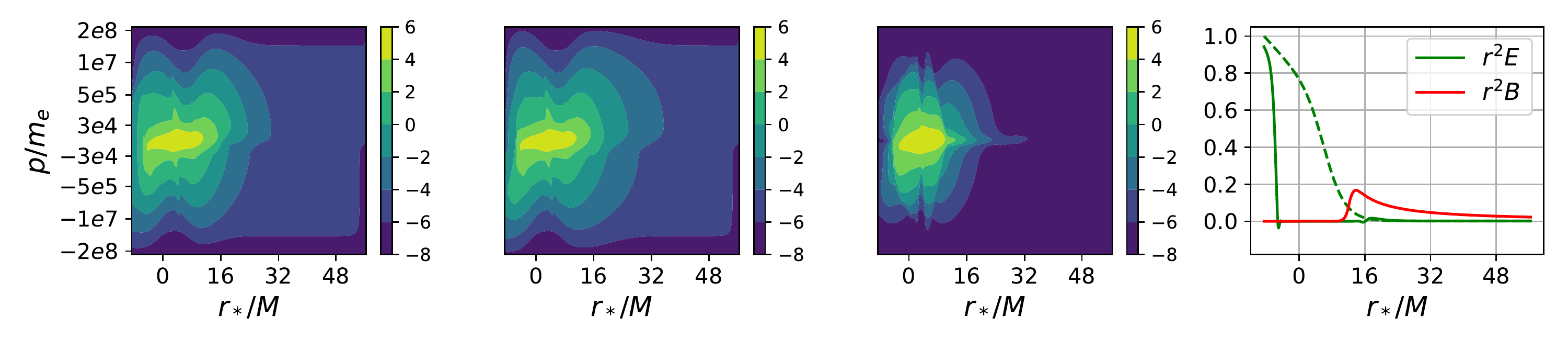}
\caption{\label{fig:evol}
BH discharge process of a fiducial model with parameters  $\Gamma_Q = 3\times 10^8$, $E_\gamma^{\rm soft} = 10$ eV,
$N_\gamma^{\rm soft} = 1$ and $\{\mathcal R_{\gamma E},\mathcal R_{\gamma B}\}= \mathcal R_{\gamma12}$. Top/Middle/Bottom rows are snapshots at $t=0.2 M/ 0.6M/ 3.7M$, respectively.
Left three columns are the distribution functions $f_{\pm,\gamma}$ (more accurately  $\log_{10}[f_{\pm,\gamma}\cdot (M\Gamma_Q m_e)]$). In the right panel, we plot EM fields $r^2E$ and $r^2B$ in unit of $M^2\times 10^{12} {\rm G}$, where the dashed lines are the initial value of $r^2E$.}
\end{figure*}

We choose a fiducial model with $\Gamma_Q = 3\times 10^8, E_\gamma^{\rm soft} = 10 \ {\rm eV},
N_\gamma^{\rm soft}=1, \{\mathcal R_{\gamma E},\mathcal R_{\gamma B} \} = \mathcal R_{\gamma12}$, and show the detailed discharge process in Fig.~\ref{fig:evol}.
In the region where electric field is not screened, electrons and positrons are
accelerated and then scattered by soft photons in a distance $1/\Ric\sim 0.1 M$.
Hard photons produced by IC scatterings are depleted by the electromagnetic fields immediately
via Schwinger process $\gamma + E/B \rightarrow e^++e^-$. As a result,
number of $e^\pm$ pairs increases exponentially while number of hard photons stays
at a relatively low level due to $\Rgg \gg \Ric$ (see 1st row of Fig.~\ref{fig:evol}).
After a few folds of exponential growth,
local number density of pairs exceeds the initial charge density, i.e., $n_+\approx n_- \gg |n_+-n_-|$,
and electric field is screened by electrons and positrons accelerated in opposite directions (see 2nd row of Fig.~\ref{fig:evol}).  As the exponential growth continues,
electric field is screened over larger and larger area, leaving a window
where EM fields vanish and hard photons survive the Schwinger process.
In the EM-free window,  leptons and hard photons are driven towards lower energy
by colliding with soft photons (see 3rd row of Fig.~\ref{fig:evol}).

\begin{figure}
\includegraphics[scale=0.65]{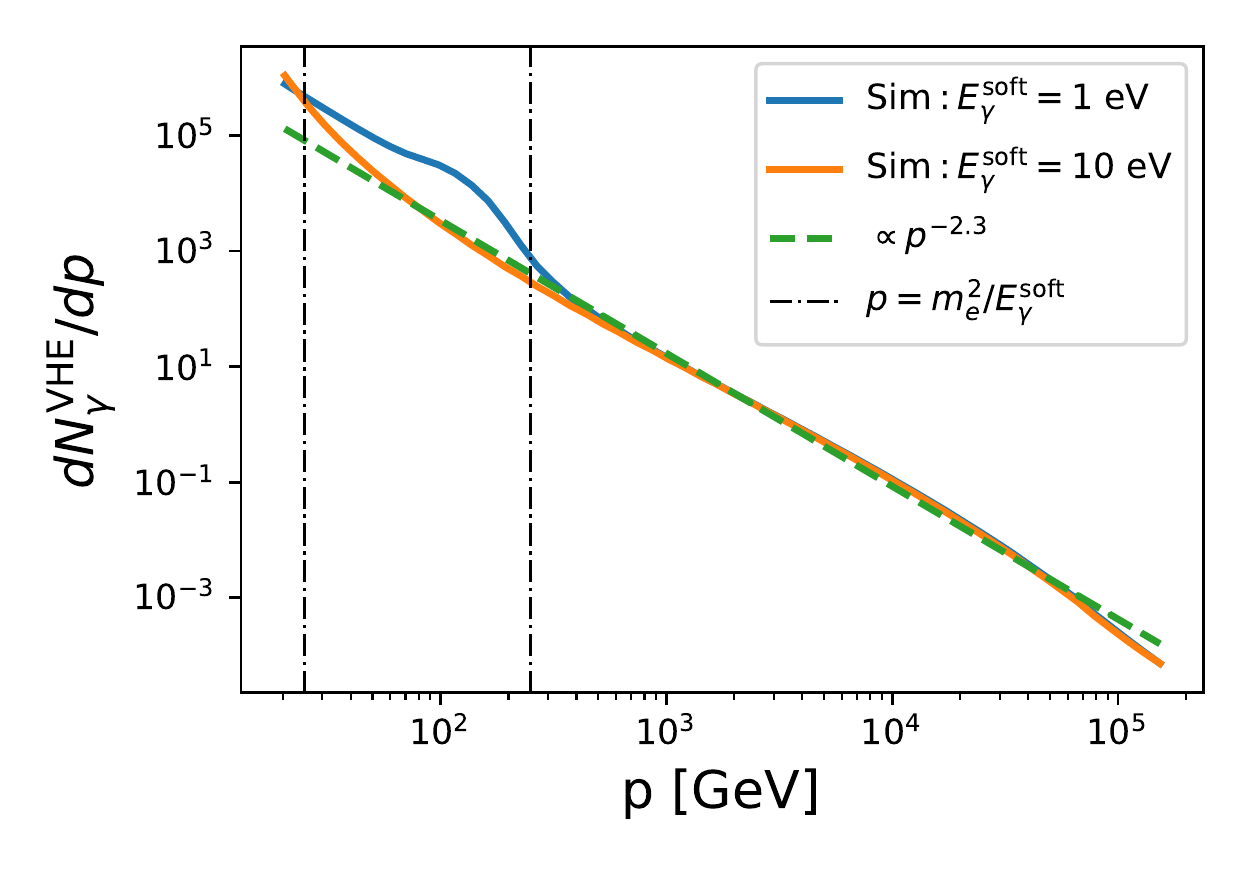}
\caption{\label{fig:sed} The energy spectrum of VHE gamma-rays observed from far-away for a model with parameters
$\Gamma_Q = 3\times 10^8$, $E_\gamma^{\rm soft} = \{1, 10\}$ eV,
$N_\gamma^{\rm soft} = 1$ and $\{\mathcal R_{\gamma E},\mathcal R_{\gamma B}\}= \mathcal R_{\gamma12}$, where
$dN_\gamma^{\rm VHE}/dp$ is given in arbitary units.
}
\end{figure}

We measure the total number of very-high-energy (VHE)  gamma rays escaping to infinity
\be
N_\gamma^{\rm VHE} = 4\pi \int_{p>50 {\rm GeV}} f_\gamma(t, r, p)|_{r\rightarrow\infty} dt dp\ ,
\ee
and the energy spectrum $dN_\gamma^{\rm VHE}/dp$. We find the energy spectrum is roughly a power law
$dN_\gamma^{\rm VHE}/dp\propto p^{-2.3}$ plus a bump at $p\approx m_e^2/E_\gamma^{\rm soft}$, which is
the energy threshold of hard photons colliding with soft photons and producing $e^\pm$ pairs (see Fig.~\ref{fig:sed}).

\begin{figure*}
\includegraphics[scale=0.7]{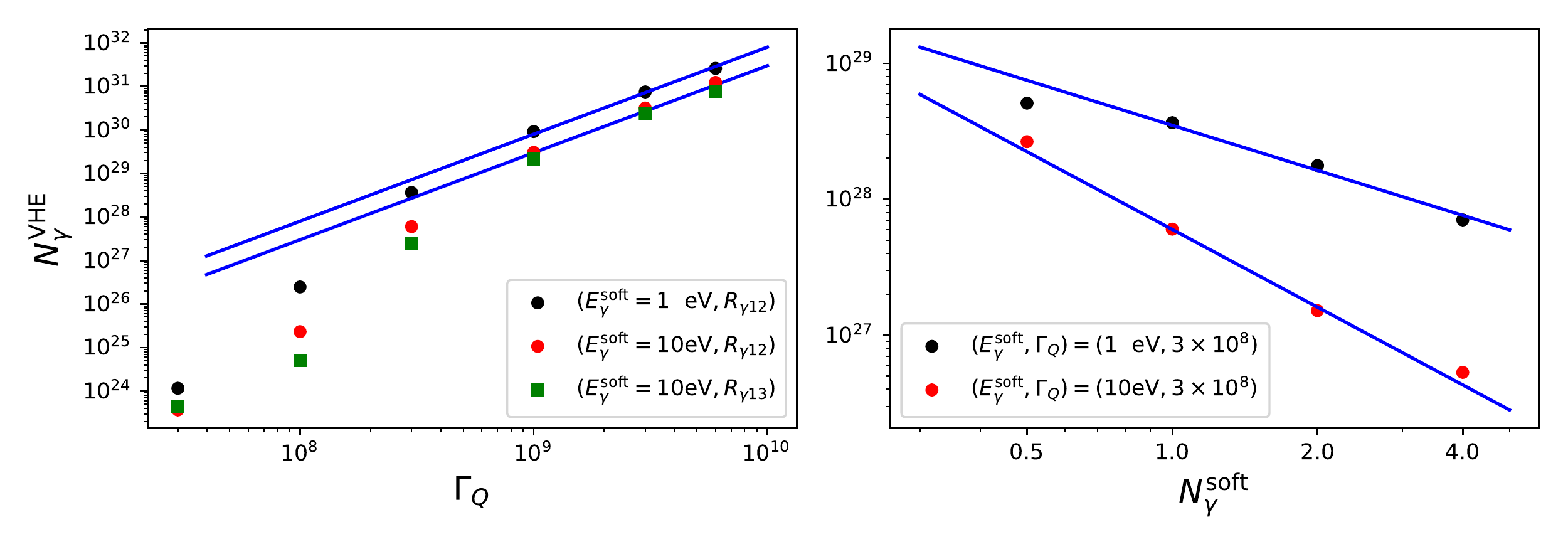}
\caption{\label{fig:Ngamma}
Dependence of VHE photon number $N_\gamma^{\rm VHE}$ on various model parameters, which take fiducial values
if not explicitly specified. In the left panel,
we show the relation of $N_\gamma^{\rm VHE}$ with initial BH charge $\Gamma_Q$, which is fitted by
$N_\gamma^{\rm VHE} = N_0\times(\Gamma_Q)^2$, with $N_0\approx 10^{12}$. In the right panel, we show
dependence of $F_\gamma^{\rm VHE}$ (fraction of initial electric energy
carried away to infinity by VHE photons) on soft photon number density $N_\gamma^{\rm soft}$,
which is roughly $F_\gamma^{\rm VHE}\propto (N_\gamma^{\rm soft})^{-\xi} $, with power index $\xi$ being $O(1)$. }
\end{figure*}

In Fig.~\ref{fig:Ngamma}, we detail the dependence of $N_\gamma^{\rm VHE}$ on the four model parameters,
which is approximately described by $N_\gamma^{\rm VHE} = N_0(E_\gamma^{\rm soft}, N_\gamma^{\rm soft}, \BNS)\times (\Gamma_Q)^2$, where $N_0|_{p>50 \ {\rm GeV}}\approx 10^{12}$ (and $N_0|_{p>20 \ {\rm GeV}}\approx 3\times10^{12}$).
The initial BH charge sets the total energy budget
($\propto\Gamma_Q^2$) for particle creation, which explains the dependence of $N_\gamma^{\rm VHE}$ on $\Gamma_Q$.
The soft photon energy $E_\gamma^{\rm soft}$ sets an threshold $ m_e^2/E_\gamma^{\rm soft}$ below which
hard photons freely stream in the soft photon background without pair creation. Therefore the smaller $E_\gamma^{\rm soft}$, the higher threshold and the more hard photons escaping to infinity. The number density of soft photons $N_\gamma^{\rm soft}$
determines the mean free path of leptons ($1/\Ric$) and that of hard photons ($1/\mathcal R_{\gamma\gamma}$), where the former determines the amount of energy a lepton
can gain from the electric field and the latter determines the annihilation possibility of a hard photon.
Therefore the larger $N_\gamma^{\rm soft}$, the less energy carried away to infinity by VHE photons.
The magnetic field strength $\BNS$ (or equivalently the coefficient $\mathcal R_{\gamma\#}$) mainly affects the width of window in which EM fields vanish and hard photons survive the Schwinger process. We expect this to be
a marginal effect which is confirmed by our simulations.

Now let us examine the detection prospect for the BH discharge process after a typical type III BH-NS merger.
For very short observation time, the sensitivity of a Cherenkov telescope is limited by
number $N_{\rm obs}$ of high-energy particles arriving at the telescope. Using the
commonly used detection criteria $N_{\rm obs} > 10$ \cite{magic2016}, we obtain
the horizon distance
\be
d =  \left( \frac{N_\gamma^{\rm VHE}}{1.2\times 10^{41}}\right)^{1/2}
\left( \frac{S}{1 \ {\rm km}^2}\right)^{1/2} {\rm Mpc}\ ,
\ee
within which BH discharge events are visible to the Cherenkov telescope,
where the effective area $S \sim 5 {\rm km}^2$ for the telescope  proposed in \cite{CTA2019}.

Our simulation is unable to trace the lowest energy photons due to limited resolution.
Assuming hard photons of energy $100$ MeV carry away total energy $5\times10^{41}$ ergs,
the horizon distance  for Fermi LAT  \cite{Fermi2009} (with effective area $\sim 0.35\ {\rm m}^2$ at this energy) is about  $0.1 \ {\rm Mpc}$.

\section{Summary and discussion}
\subsection{Summary}
In this
paper, we perform a first-principle study of EM counterparts of type III BH/NS mergers, where the BH
is large enough to swallow the NS as a whole without star disruption. In the late inspiral stage,
the BH is charged as moving in the NS magnetosphere. The  remnant BH
neutralizes rapidly via a discharge process, through which the electric energy is converted into the kinetic energy
of $e^+/e^-$ pairs and hard photons. For a typical BH/NS merger with $M=10 M_\odot$, $\BNS = 10^{12}$ G and $r_{\rm NS} = 10$ km,
we find that $\sim 10\%$ of the total electric energy is
carried away to infinity in a time interval $\sim 1$ ms by VHE gamma rays,
whose  spectrum is approximately a power law $dN_\gamma^{\rm VHE}/dp\propto p^{-2.3}$ plus a bump
at $p\approx m_e^2/E_\gamma^{\rm soft}$ (see Fig.~\ref{fig:sed}).

In reality, soft photons from cyclotron radiation in the non-uniform magnetic field are not mono-energetic.
As shown in Figs.~\ref{fig:sed} and \ref{fig:Ngamma}, different soft photon energy indeed mildly changes the total number
and the energy spectrum of VHE photons. We expect a mild increase in the total number $N_\gamma^{\rm VHE}$ and flatter
bump in the spectrum $dN_\gamma^{\rm VHE}/dp$ if the low-energy tail of soft photons are taken account of.

We may similarly estimate  the EM signal from a BH/magnetar merger. For a magnetar with magnetic field $10^{15}$ G \cite{Woods:2003si,Mereghetti:2015asa},  the final BH would be charged to $10^3 Q_{,12}$ with  electric energy $\sim 10^{48}$ ergs. Soft photon energy ($\hbar\omega_B$) would be $10^3$ larger, and typical hard photon energy $m_e^2/E_\gamma^{\rm soft}\sim 25$ MeV.
Therefore we expect a dim sGRB after a BH/magnetar merger which is visible to Fermi LAT within $\sim 100$ Mpc.
The discharging process discussed here  also sheds light on the EM signature of charged BH mergers \cite{zhang2016mergers,liu2016fast} and/or  charged BHs formed through gravitational collapse of magnetized NSs \cite{Nathanail:2017wly}.

\subsection{Discussion}
It is beneficial to compare the BH discharge process with that of a NS. As shown in \cite[e.g., Ref.][]{Zhang2019},
a spinning NS carrying a magnetosphere is also charged with electric charge $Q_{\rm NS}\sim \Omega_{\rm NS} B_{\rm NS} r_{\rm NS}^3$,
where $\Omega_{\rm NS}$ is the NS angular velocity. The NS magnetosphere is approximately being force-free
(the component of the electric field parallel to the magnetic field $E_\parallel$ vanishes), except
in some special regions, e.g., the polar caps. Inside the polar cap, a cascade of pair creation and
synchrotron radiation is driven by the strong
unscreened electric field $E_\parallel$ \cite{Daugherty1982,Zhang2000}. As a result, electrons and positrons are accelerated
by the parallel electric field $E_\parallel$ in opposite directions. This process is very similar to the BH discharge
studied in this paper.
However, the cascade in the NS polar caps does not continually decrease the NS charge $Q_{\rm NS}$ due to the existence of a
recovery force $B_{\rm NS}$ which is unaffected by the cascade and drive a electric current circuit along open magnetic field lines keeping both the electric field and the NS charge in
the equilibrium state. To summarize, it is the NS magnetic field $B_{\rm NS}$ that keeps the NS from being discharged.

An essential difference between a (isolated) BH and a NS is that the BH cannot sustain the magnetic field captured from e.g., a BH-NS merger. Unlike NSs which are almost perfect conductors capable of trapping magnetic fields, BHs are dissipative according to the membrane paradigm \cite{Thorne1986}. Therefore we do not expect long-lived magnetic fields around black holes, if there is no external driving source.
If a cascade of pair creation and high-energy photon production is ignited, the initial BH charge $Q_{\rm BH}$ will inevitably deplete because there is no such recovery force that plays the role of replenishing current circuit as in the spinning NS case. Both electric and magnetic fields decrease monotonically as the BH charge decreases.

Of course, the BH discharge picture will be completely different if we are considering a spinning BH immersed in a magnetic field $B_{\rm ext}$ sourced by some \emph{external electric current},
e.g., sustained by a conducting accretion flow. In this case, the BH will not be discharged by the cascade as the magnetic field $B_{\rm ext}$
now serves as the recovery force maintaining both the BH and the magnetosphere in the charged state.

Throughout this work, we have assumed spherical symmetry for the system of charged BH and surrounding particles,
and have solved the  Boltzmann equations of $1+1$ form. The key elements of the BH discharge process
are the initial BH charge (energy budget) and the remnant NS magnetosphere (soft
photon source). For a Kerr-Newman BH, there will be both electric and magnetic fields measured by,
e.g., Carter observers \cite{Carter1968}.  Both fields ($E\sim Q/r^2$ and $B\sim 2aMQ/r^3$ with $a$ being the dimensionless BH spin \cite{Znajek1977}) are proportional to the BH charge and will be screened by the leptons in a similar way
as in the non-spinning BH case. In a naive picture, the Kerr-Newman BH is approximately
a charged spinning object of size $\sim M$ and carrying
both net charge and a current loop, which source the E-field and the B-field, respectively. With the ignition of pair production,
particles of opposite charge are attracted towards the central BH until the total charge is neutralized and therefore the E-field is screened. While the charged particles
move inward, they are forced to rotate with the B-field lines and generate an opposite current loop until the initial current loop is
counteracted and therefore the B-field is screened. A non-zero BH spin does not introduce any recovery force as in the NS case, therefore we expect the discharge process also to be true for spinning charged BHs.
To understand the details of 3D discharge process for spinning BHs where electrodynamics of higher dimensions kick in,
our low-dimension analysis does not work any more and it is preferable to apply the particle-in-cell simulations \cite{Levinson:2018arx,Parfrey:2018dnc,Chen:2018khs}.

\begin{acknowledgements}
We would like to thank the referees for their comments which enable a substantial improvement of this paper.
We thank Junwu Huang and Luis Lehner for very helpful discussions.
This research was also supported by  the Natural Sciences and Engineering Research Council of Canada and in part by Perimeter Institute for Theoretical Physics.
Research at Perimeter Institute is supported in part by the Government of Canada through the Department of Innovation, Science and Economic Development Canada and by the Province of Ontario through the Ministry of Economic Development, Job Creation and Trade.
\end{acknowledgements}

\appendix*
\section{Sanity Check}
\begin{figure*}
\includegraphics[scale=0.7]{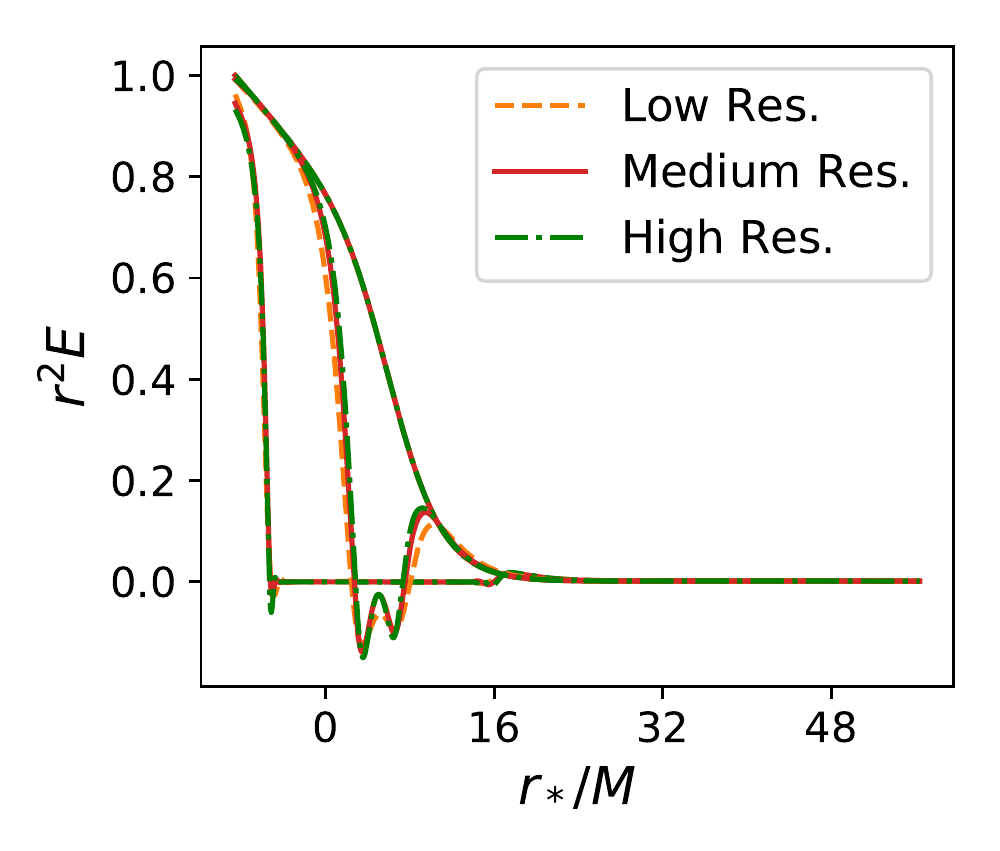}
\includegraphics[scale=0.7]{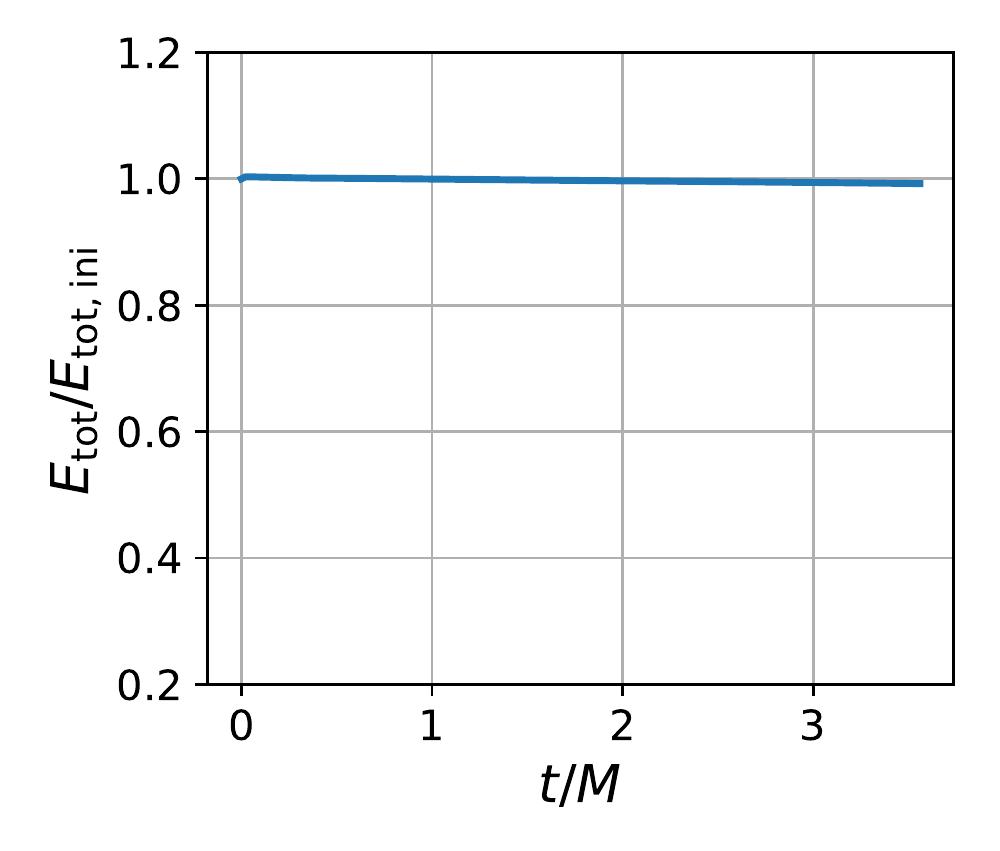}
\caption{\label{fig:check}
In the left panel, we show
the discharge process at time $t=0/0.6/3.7 M$ for three different resolutions. In the right panel, we show the precision
of energy conservation for the simulation with the default medium resolution, where the total energy $E_{\rm tot}$ is the total energy
of the whole system, including energy of the electric field, and energy of high-energy pairs and photons (residing in the computation
domain and escaping to infinity).}
\end{figure*}
As a sanity check of our simulations, we run a convergence test and a conservation test for the fiducial model
(see Fig.~\ref{fig:evol} for the model parameters).
For the convergence test, we evolve the governing equations with three different resolutions and find the simulations
are well converged for the default medium resolution we used throughout this paper (see Fig.~\ref{fig:check} for a glimpse).
For the conservation test, we keep track of the total energy $E_{\rm tot}$ of the whole system, including energy of the
electric field, energy of all particles residing in the computation domain and energy of particles captured by the central BH or escaping
to infinity. We find the energy conservation is satisfied with percent-level precision.

\bibliography{ms}

\end{document}